\documentclass[a4paper,12pt,floats]{revtex4-1}

\usepackage{graphicx,epsfig}
\usepackage{times}
\usepackage{amssymb,amsmath}

\newcommand{\etal}{\emph{et al.} }

\begin{document}

\title{Urban street networks: a comparative analysis of ten European cities}

\author{Emanuele Strano}
\affiliation{Laboratory of Geographic Information Systems (LASIG), School of Architecture, Civil and Environmental Engineering (ENAC), Ecole Polytechnique F{\'e}d{\'e}rale de Lausanne (EPFL)} 
\affiliation{UDSU, Urban Design Studies Unit, Department of Architecture, University of Starthclyde, Glasgow, UK}

\author{Matheus Viana}
\affiliation{Institute of Physics of Sao Carlos, University of Sao Paulo, Sao Paulo, Brazil} 

\author{Alessio Cardillo} 
\affiliation{Departamento de F\'{\i}sica de Materia Condensada, Universidad de Zaragoza, E-50009 Zaragoza, Spain}
\affiliation{Institute for Biocomputation and Physics of Complex Systems (BIFI), Universidad de Zaragoza, E-50018 Zaragoza, Spain}
\affiliation{Dipartimento di Fisica e Astronomia, Universit\`a di Catania and INFN, Via S. Sofia, 64, 95123 Catania, Italy} 

\author{Luciano Da Fontoura Costa}
\affiliation{Institute of Physics of Sao Carlos, University of Sao Paulo, Sao Paulo, Brazil} 

\author{Sergio Porta}
\affiliation{UDSU, Urban Design Studies Unit, Department of Architecture, University of Starthclyde, Glasgow, UK}

\author{Vito Latora}
\affiliation{School of Mathematical Sciences, Queen Mary, University of London, London, UK}
\affiliation{Dipartimento di Fisica e Astronomia, Universit\`a di Catania and INFN, Via S. Sofia, 64, 95123 Catania, Italy} 
\affiliation{Laboratorio sui Sistemi Complessi, Scuola Superiore di Catania, Via San Nullo 5/i, 95123 Catania, Italy}

\date{\today}

\begin{abstract}
We compare the structural properties of the street networks of ten different European cities
using their primal representation. We investigate the properties of the geometry of the networks
and a set of centrality measures highlighting differences and similarities among cases.
In particular, we found that cities share structural similarities due to their quasi planarity
but that there are also several distinctive geometrical proprieties. A Principal Component
Analysis is also performed on the distributions of centralities and their respective moments,
which is used to find distinctive characteristics by which we can classify cities into families.
We believe that, beyond the improvement of the empirical knowledge on streets network
proprieties, our findings can open new perspectives in the scientific relation between city
planning and complex networks, stimulating the debate on the effectiveness of the set of
knowledge that statistical physics can contribute for city planning and urban morphology
studies.\\
\newline
\textbf{keywords:} complex street networks, urban form, city classification, centrality.
\end{abstract}


\maketitle

\newpage

\section{Introduction}

Defining urban form is certainly an important and difficult issue, especially if one wants to supply useful knowledge to urban planners and urban designers or new knowledge for city scientist.
In this paper we address this question, and we try to improve the empirical-based knowledge on the structure of a city by studying the urban street networks of ten European cities, namely Edinburgh, Leicester, Sheffield, Oxford, Worcester, Lancaster, Catania, Barcelona, Bologna and Geneva.
The form of cities is the subject of an area of urban studies named urban morphology. Urban morphology in its current form emerged between the 40s and the 60s of the XX Century from the work of two scholars as prominent as different: the German-born and then British urban geographer M.R.Conzen (1960), and the Italian architect and historian Saverio Muratori (1960). In this area, the main subject of investigation is the urban fabric of the city at the scale of the neighborhood, street, plot and building. 

A different branch of urban morphology has stemmed from the sciences of complex systems building on a long standing tradition in regional analysis, economic geography and modeling \cite{Anas1998}. Complexity in the built environment is here investigated with the same instruments used for other classes of self-organized phenomena in nature, technology and society \cite{BattyBook}. These works are now flanked by a growing interest on complex spatial networks within the community of physics \cite{Vito2006}.

Spatial networks and in particular planar graphs are suited to model a number of real phenomena \cite{Barthelemy2011}.  We are here interested in the study of a particular class of spatial networks that describe the street pattern of cities. The beginning of these studies can be traced back to the classical works on regional transportation networks based on graph theory \cite{Garrison1962, Kansky1989}.  The advent of complex system science and its paradigm \cite{Barabasi1999, Barabasi2002}, jointly with the increasing availability of spatial and time geo-referenced data, has given a new boost to these studies, and several important contributions have appeared recently \cite{Barthelemy2011, Strano2012}. 

In \cite{Masucci2009}, Masucci \etal study the structural property of the London street network in its dual and primal representation.  In \cite{Jiang2007}, the autors, by using 40 urban networks in a dual representation, found a small-world structure and a scale-free property for both street length and connectivity degree, and used various centrality indices as indicators of the importance of streets.  L\"ammer \etal developed a comparative analysis of the betweenness distribution in 20 cities in Germany, suggesting a relation with vehicular traffic \cite{Lammer2006}.  Others have focused on centralities in primal and dual representations of street networks \cite{Porta2006B, Porta2006, Porta2006B, Crucitti2006,Hillier1984,Hillier1999}, and on other structural features, such as the number of cycles of a given length \cite{Cardillo2006}.

However, an important and still open question in urban morphology has to do with the characterization of classes of cities based on their form. This is a preliminary step to approach a comparative analysis aimed at the classification of cities. In this paper we propose a classification based on the distribution of street centrality by cross-comparing real cases and therefore with no use of null models. We limit our study to the characterization of city form without exploring its impact on collective behaviors, an area of research which, at the scale of entire cities, is now finding new opportunities through the exploitation of massive datasets from online geo-social networks and mobile geo-referenced systems \cite{Rattiplos2010, Expert2011}

In our study, we first observe the geometry of the networks following the approach recently suggested in \cite{Chan2011}; we consider the distribution of basic indices of the primal street networks such as street angles at intersections and street lengths. Secondly, we investigate four different centralities indices computed over the entire urban street network, highlighting their distribution and their mutual correlation. 

We  considered 10 European cities namely  Edinburgh, Leicester, Sheffield, Oxford,
Worcester, Lancaster, Catania, Barcelona, Bologna and Geneva. We show that these cities share some structural geometric properties, which are the same found in other planar spatial systems such as those of leaves \cite{Perna2011,Couder2002, Bohn2002} , which suggests that planarity in itself is a driver across spatial systems in various domains.

However, we also show that cities are different. We witness that some cities, like Catania, stand alone in terms of basic geometric properties and, more importantly, that the various distributions of centrality are reconcilable to one common pattern (a power law) only if a largely minoritarian subset of streets are taken into consideration. These results have to do with the extreme heterogeneity of the cities' visible traits as resulting from the interplay of entirely different phenomenon in time, such as historical accidents (including planning), physical constraints or just random events \cite{BattyBook}. In particular, we show how cases tend to cluster in groups after the whole set of centrality measures distributions are considered in a way that suggests major planning interventions and physical geographic constraints are key.

The paper is organized as follows. In Section~\ref{s:basicprops} the case studies are presented and their geometric properties are introduced and analyzed. In Section~\ref{s:centralities} the study of the four centrality indices is illustrated along with the clusterization of cases according to their combined behavior as a result of the application of a Principal Component Analysis (PCA) of the distribution of those four centralities. Finally in Section ~\ref{s:discussion} we offer a discussion of the results and our conclusions.

\section{Basic proprieties of urban networks}
\label{s:basicprops}

\begin{figure}[hb!]
\centering
\includegraphics [width=6cm] {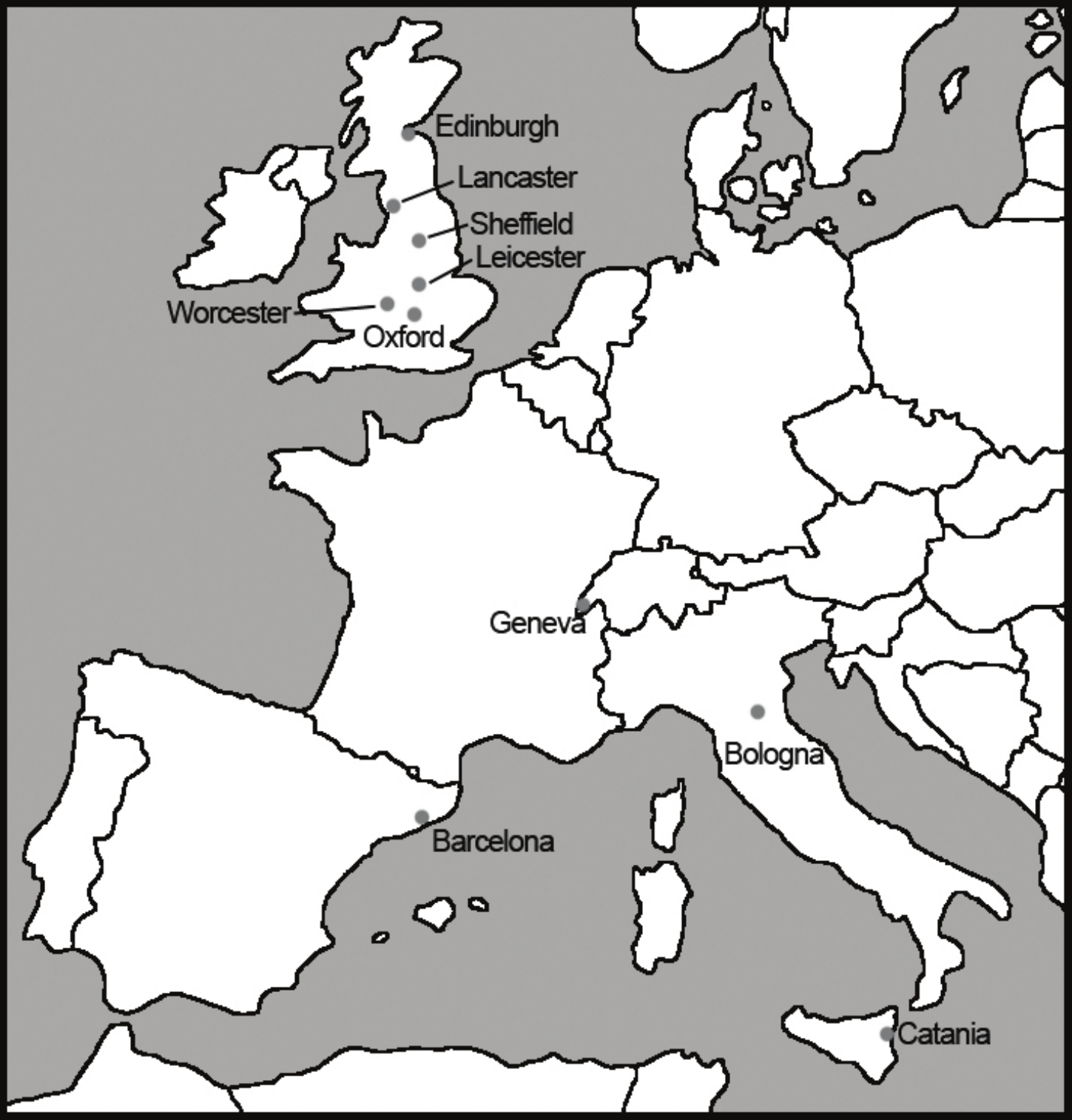}	
\caption{Geographical location of the cities.}
\label{f:map}
\end{figure}

We address the analysis of 10 European cities, namely: Edinburgh, Leicester, Sheffield, Oxford,
Worcester and Lancaster (UK); Catania and Bologna (IT); Barcelona (ES); Geneva (CH) (Fig.~\ref{f:map}). These cities are variously located and present remarkably different economic, cultural and climatic conditions along with great variety of characteristics like population and size (Tab.~\ref{t:ppts}). We represent the street network of these cities such that intersections are nodes and streets are links between nodes, i.e. a primal representation \cite{Porta2006}. We analyse such networks in terms of their basic properties and several geometric indices such as street length and the angle they form at intersections.

Before getting deeper into the analysis, we introduce some basic concepts of graph theory. A graph (or network) is a mathematical object which consists of two sets: $\mathcal N$ and $\mathcal L$. The $N$ elements of the former are called \emph{nodes}, while the $E$ elements of the latter (unordered couple of nodes) are called \emph{links}.\\ There are many ways to represent a graph, but, the most common one is the \emph{adjacency matrix} $\cal A$, a $N \times N$ square matrix whose entry $a_{ij} \, (i,j = 1 , \ldots , N )$ is equal to one if link between nodes $i$ and $j$ exists and zero otherwise. The \emph{degree} of a node $i$, $k_i$, is the number of links incident with it. The \emph{average degree} $\langle k \rangle = 2E/N$, is nothing but the average of the degrees over all the nodes in the network.

\begin{figure*}[hc!]
\centering
\includegraphics [width=15cm] {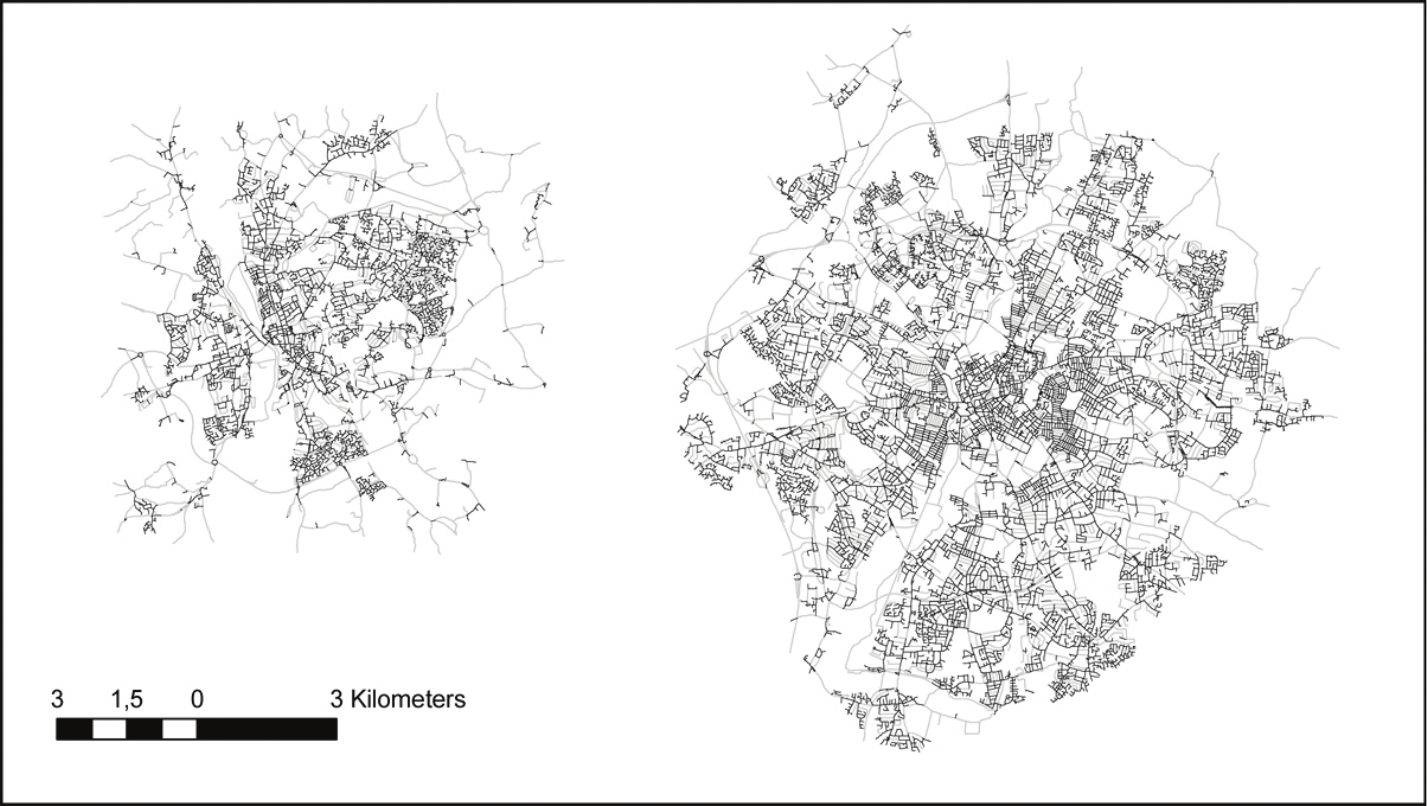}	
\caption{Example of street networks, Leicester (Left) and Worcester (Rigth). The streets with length that are not following a power-law distribution are shown in black, while streets with length 
that follows a power-law distribution are shown in grey.}
\label{f:nets}
\end{figure*}

Networks of street patterns belong to a particular class of graphs called, \emph{planar graph}, i.e. graphs whose links cross only on nodes. In our case, the nodes represent street intersections, while the links are the streets centerline, a network made using this convention is called \emph{primal network} (Fig.~\ref{f:nets}). Our networks are also weighted and each link ($i$,$j$) carries a numerical value $w_{ij}$ expressing the intensity of the connection. The natural choice, in our case, for the functional form expressing the weight of a link connecting nodes, say $i$ and $j$, is to put $w_{ij}$ equal to the length of the street connecting, $l_{ij}$.

Our analysis starts by importing the street system into a Geographical Information System (GIS) environment. Data of the street systems have been retrieved from different sources: for example, in all UK cases we have used the Ordnance Survey maps, while in Italian cases we have used data from City Councils planning offices, and in Barcelona we worked on a dataset provided by the Ag\`encia d'Ecologia Urbana de Barcelona. Given that these geographical street networks had been mainly built for the sake of traffic navigation or planning, they presented characteristics that not always fitted the purposes of a centrality analysis; for example, multi-lane streets were usually represented with one link per lane rather than one link per street. As a result we have prepared our databases by first cleaning the networks accurately to remove link redundancies, fix short missing links, collapse unconnected links on the same node when needed, continuously confronting the networks with aerial images of the real cities drawn from remote sensing sources such as Google Earth. Such procedure was undertaken both manually and through ad-hoc tools in a GIS environment. For the definition of the boundary of the urban systems, we followed the border of the built-up area extending it by roughly 1km outbound.

Considering the entries of Tab.~\ref{t:ppts}, we can see how various selected cities are for example in terms of size, from small cases like Lancaster to large ones like Barcelona, or in terms of street intersection density, from very dense cities like Catania to more sprawled ones like Edinburgh. We have selected cities with different levels of geographic constraints, from those like Geneva and Oxford traversed or limited by large natural water features to those like Catania and Bologna that sit on uninterrupted plains, and a different prominence of planning history, from those self-organized or only fragmentarily planned like Leicester or Bologna to one like Barcelona whose street layout had been heavily determined by one single planning vision (the 1859 Cerd\'a Plan). We see that the particular planning history of Barcelona is reflected in the low values of both the standard deviation of the street length and the percentage of dead-end streets on the total number of streets, both resulting from the extensive adoption of a rigid homogeneous grid layout. The extreme diversity of selected cases has been pursued in order to make the comparative analysis of similarities and differences more profound.

\begin{table*}[ht!]
\centering
\begin{tabular}{|l|c|c|c|c|c|c|c|c|c|c|}
\hline
\textsc{City} & \textsc{Population} & \textsc{Area} (km$^2$) & $N$ & $E$ & $\langle k \rangle$ & $\rho$ (km$^{-1}$) & $L$ (km) & $\langle \ell \rangle$ (m) & $\sigma_{\ell}$ (m) & $f$ (\%)\\
\hline
Barcelona & 1,615,908 & 82.0 & 6452 & 11071 & 3.43 & 15.15 & 1242 & 110.7 & 105.1 & 0.1 \\
Bologna & 380,878 & 88.6 & 5200 & 7359 & 2.84 & 9.19 & 814 & 119.1 & 158.0 & 23 \\
Catania & 293,811 & 34.0 & 11099 & 14039 & 2.52 & 23.91 & 813 & 55.9 & 57.3 & 17 \\
Edinburgh & 477,660 & 195.6 & 5021 & 13063 & 2.43 & 8.96 & 1752 & 110.0 & 147.0 & 24\\
Geneva & 191,237 & 95.0 & 6183 & 8681 & 2.80 & 11.18 & 1062 & 122.4 & 119.7 & 0.9 \\
Lancaster & 45,952 & 77.7 & 5913 & 15567 & 2.51 & 9.28 & 721 & 96.8 & 153.6 & 18 \\
Leicester & 288,000 & 122.2 & 7186 & 8896 & 2.47 & 7.23 & 883 & 98.5 & 94.5 & 18 \\
Oxford & 149,800 & 51.1 & 4372 & 11071 & 2.32 & 10.27 & 525 & 103.0 & 133.7 & 28  \\
Sheffield & 520,700 & 187.5 & 14583 & 17674 & 2.42 & 10.58 & 1983 & 111.0 & 129.8 & 21 \\
Worcester & 94,300 & 45.1 & 4685 & 5538 & 2.36 & 11.75 & 530 & 94.8 & 126.6 & 21 \\
\hline
\end{tabular}
\caption{Basic proprieties of the primal networks. $N$ and $E$ are the number of nodes and number of edges, respectively, while $\langle k\rangle$ is the average degree. The density $\rho$ is given by the ratio between the total length ($L$) and the Area. $\langle \ell \rangle$ indicates the the average length of the edges, while and $\sigma_{\ell}$ corresponds to its standard deviation. $f$ indicate the percentage of tree-like appendixes.}
\label{t:ppts}
\end{table*}

The study of the geometric properties of the networks has been focused on the distribution of three indices with the aim to find common patterns: street length, angles formed between street intersections, and the relation between dead-end link length and the area of the cycle they belong to. Following the definition given in \cite{Chan2011}, we consider \emph{cycles} as polygons formed by closed loops of links. We build our approach on previous findings that have identified universal geometric patterns under seemingly diverse street networks in cities \cite{Masucci2009,Barthelemy2008,Lammer2006,Perna2011,Couder2002, Bohn2002} and extend our exploration to focus on local patterns that actually make for the uniqueness of each case or clusters of cases.

%
%
In a city both long and short streets play an important role. The former allow connecting distant locations, the latter act as shortcuts between longer streets reducing the average path length in the navigation of the system. Short and long streets have a different historical meaning in the evolution of cities, as street length tends to diminish with increasing density of the urban pattern, following a rule according to which the more ``urban'' the area the shorter the street. 

The overall average street length $\langle \ell \rangle$ of selected networks represents a simple and good indicator of the diversity of cities, and looking at Tab.~\ref{t:ppts} we find a considerable variation of such average that, even putting aside the special case of Catania with $\langle \ell \rangle$ = 56m, which is due to the extreme density of the historical core, lays between 94.8m (Worcester) and 122.4m (Geneva).

We have then observed the distribution of street length in the selected cases. Since we are comparing cities with different size, we have considered the normalized length $\ell$, i.e. the street length divided by the diameter of the network defined as the maximum Euclidean distance between any couple of nodes belonging to the network.

\begin{figure}[ht!]
\centering
\includegraphics [width=7cm] {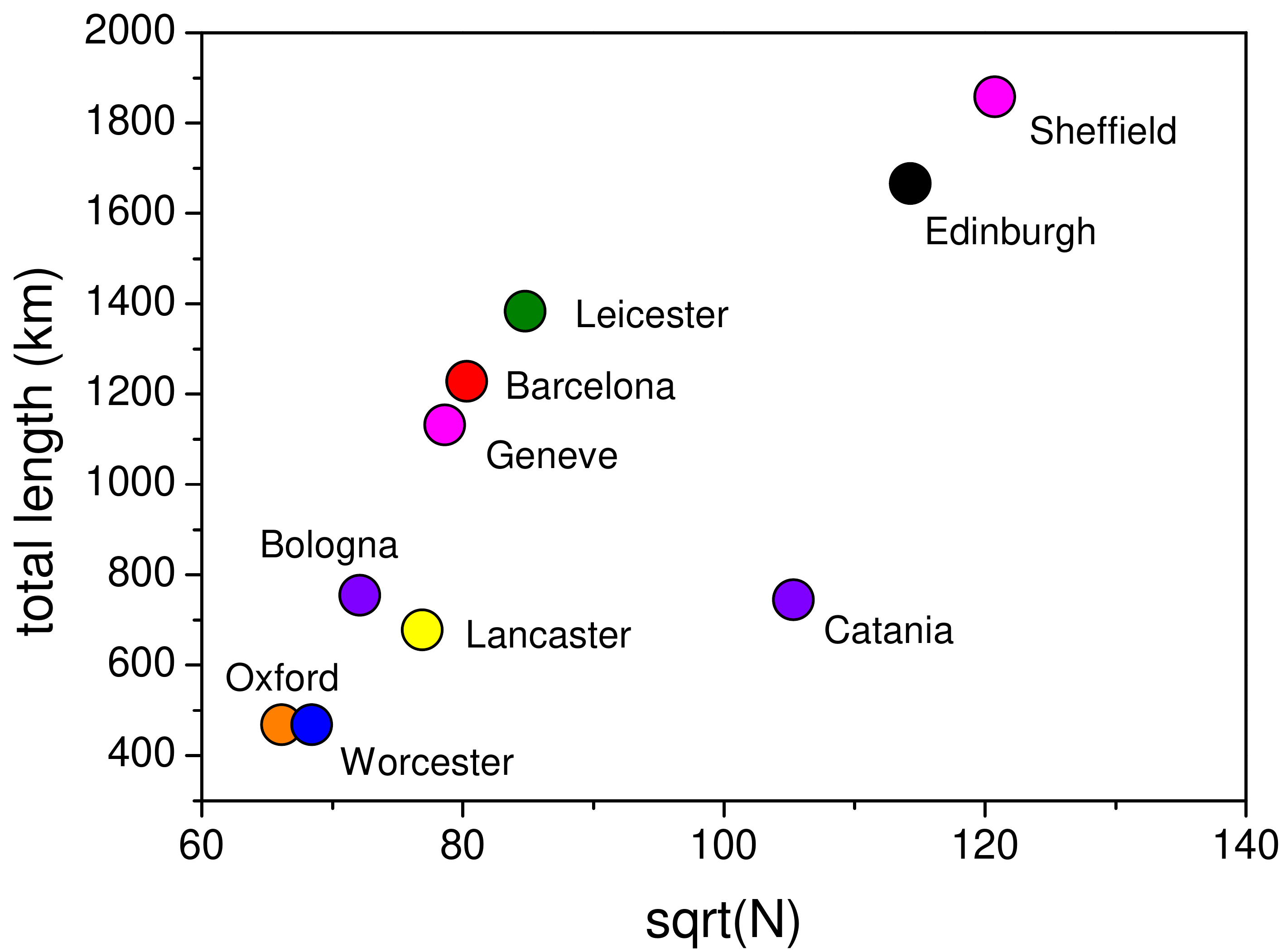}
\caption{Total street length $L$ as a function of $\sqrt{N}$.}
\label{f:lengthnodes}
\end{figure}

In Fig.~\ref{f:lengthnodes} we confirm the findings of \cite{Barthelemy2008}, that the relation between the total length $L$ and the number of nodes $N$ scale as $\sqrt{N}$. However, we notice that our cases are not closely distributed along a straight line, indicating a significance variance that can be explained by the different nature of our dataset: we are in fact comparing a smaller number of cases; moreover, our cases are large networks representing entire cities, which means that we are here handling non homogeneous and invariant street networks made of parts derived by different historical formations and shape. It is exactly this variability that we want to capture with a closer look at the differences emerging from the data.

\begin{figure*}[hc!]
\centering
\includegraphics [width=15cm] {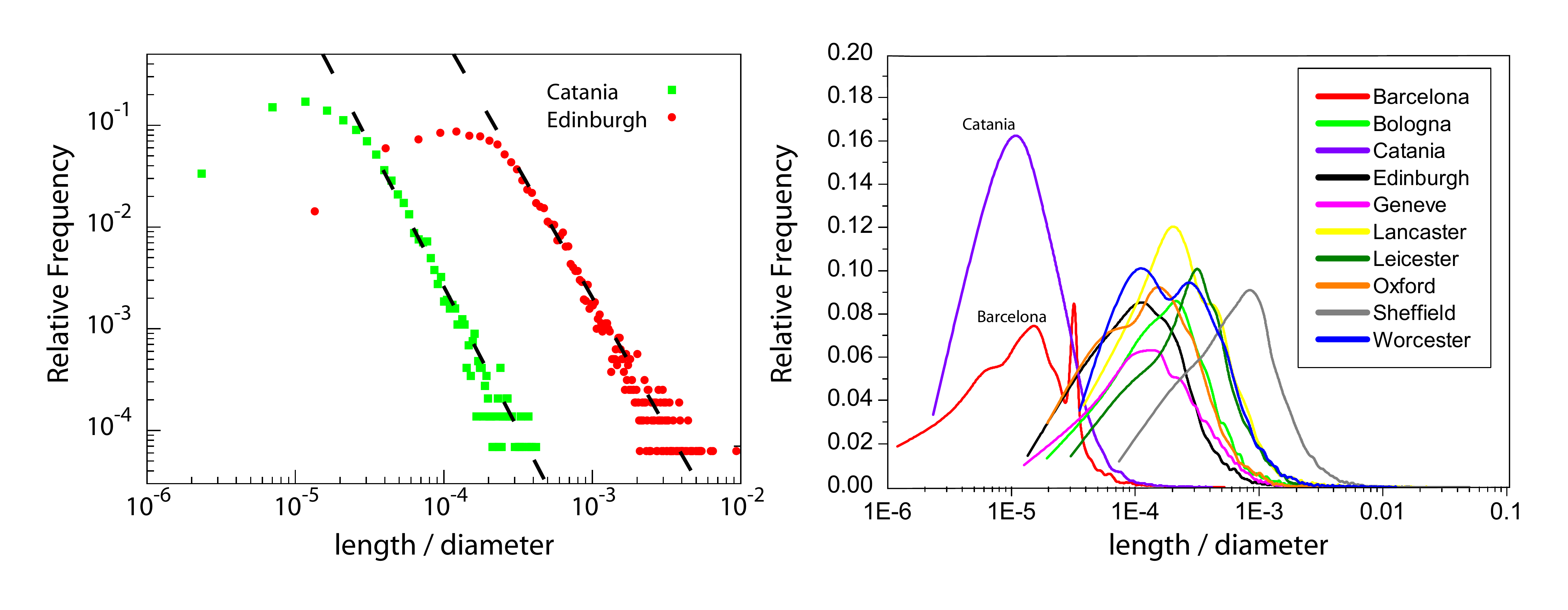}	
\caption{Distribution of normalized length $\ell$ for the ten cities in semi-log scale (right) and for two representative cities, Catania and Edinburgh, in the log-log scale (left). It is possible to observe how different visualization give different results. While in the semi-log plot the distribution of the majority of streets seem different, in the log-log they have the same trend.}
\label{f:lengthdis}
\end{figure*}

The distribution of streets' length is the simplest and first indicator that we use for changing such differences. Previous findings proposed a power law distribution with a cut-off for the longest street segments \cite{Masucci2009} and a bimodal shape distribution with a plato region above 30m \cite{Chan2011}. We are finding here slightly different results. If we look at Fig.~\ref{f:lengthdis}a, we see that a power law emerges in the distribution tail though, of course, the accordance between the distribution and the fitting became worse with decreasing of the streets' length. However, since we want to look at local patterns as well as global, we want to investigate what happens in the region that is not well fitted by a power law. 
 
In Fig.~\ref{f:lengthdis}b, we plot also the same distribution in a semi-log scale.  Here it is possible to see that most of the distributions exhibit a peak in the region between $8\times 10^{-4}$ and $3\times 10^{-3}$, which means that the majority of streets have a normalized length around that values. However Catania, Barcelona and Worcester have different behaviors whose causes might be traced back to historical accidents. Catania falls out of these boundaries because of its abundance of very short streets in the historical urban center, possibly a consequence of the complete reconstruction of the city center the disastrous eruption of the Etna Volcano in 1669. Barcelona presents a anomalous peak clearly related to the Plan Cerd\'a, mentioned above, a massive grid-iron plan covering the central part of the current urban area. Worcester exhibits a double peak in the length distribution, a consequence of the post XXII planning process that clearly shapes the most of the periphery. From this simple recollection we can appreciate the impact of specific historical occurrences that mostly have impacted on the urban form in terms of an interplay between planned and the non-planned urban forms.

This variety of street patterns gets entirely hidden in the conventional representation of data through log-log charts and if we look at what part of the street layout falls within the region that is correctly fitted by a power law function Fig.~\ref{f:lengthdis}a, we see that it actually represents a vast minority of the entire network in all our cases. 

In terms of sheer number of streets, using the method proposed in \cite{Clauset2009} we observed that the percentage of streets falling inside the power law region ranges from 4\% in Barcelona to the 29\% in Lancaster as shown in Table ~\ref{t:pl}. Of course, streets falling in the power law region are the longest, so they cover a larger share of the system in terms of street length; however, even so, the percentage of the total street length belonging to streets falling outside the power law region in most of cities it is up to the 60\% as shown in the last column of Table~\ref{t:pl}. We can appreciate visually the geographic consistency and character of this left over from power law portion of the street network in Leicester and Worcester in the Fig.~\ref{f:nets}: clearly, this portion represents not only the majority of the street network, but also the part that is historically more important, the denser and the more central, which is no surprise if we think that it is made by the shortest streets however generalization and conclusions may be improved by further and deeper tests on larger areas. For the time beeing, based on this resutls, we may argue that cities are composed by streets following two distribution and that it may reflect different dynamics of urban evolution. 

\begin{center}
\begin{table*}[!ht]
\centering
\begin{tabular}{|l|c|c|c|}
\hline
\textsc{City} & Threshold (m) & \% of streets in a Power Law distribution & \% of $L$ not in the Power Law\\
\hline
Barcelona  & 347 &   4 &   86\\
Bologna &  122 &   28  &  37\\
Catania &  143  &  6  &  77\\
Edinburgh &  194 &   13  &  59\\
Geneve  &  218  &  13  &  63\\
Lancaster  &  93 &   29 &   32\\
Leicester &  211  &  9  &  74\\
Milano  &  233 &   8  &  71\\
Oxford  &  184  &  13  &  58\\
Sheffield &  245 &   8  &  70\\
Worcester &  166 &   14 &   55\\
\hline
\end{tabular}
\caption{The total amount of streets are following a Power Law distribution is very low and includes only the longer streets of the city. In terms of total street length the percentage of street are not following the Power Law distribution are always the majority with the exception of Lancaster.}
\label{t:pl}
\end{table*}
\end{center}
%

\begin{figure}[hc!]
\centering
\includegraphics [width=7cm] {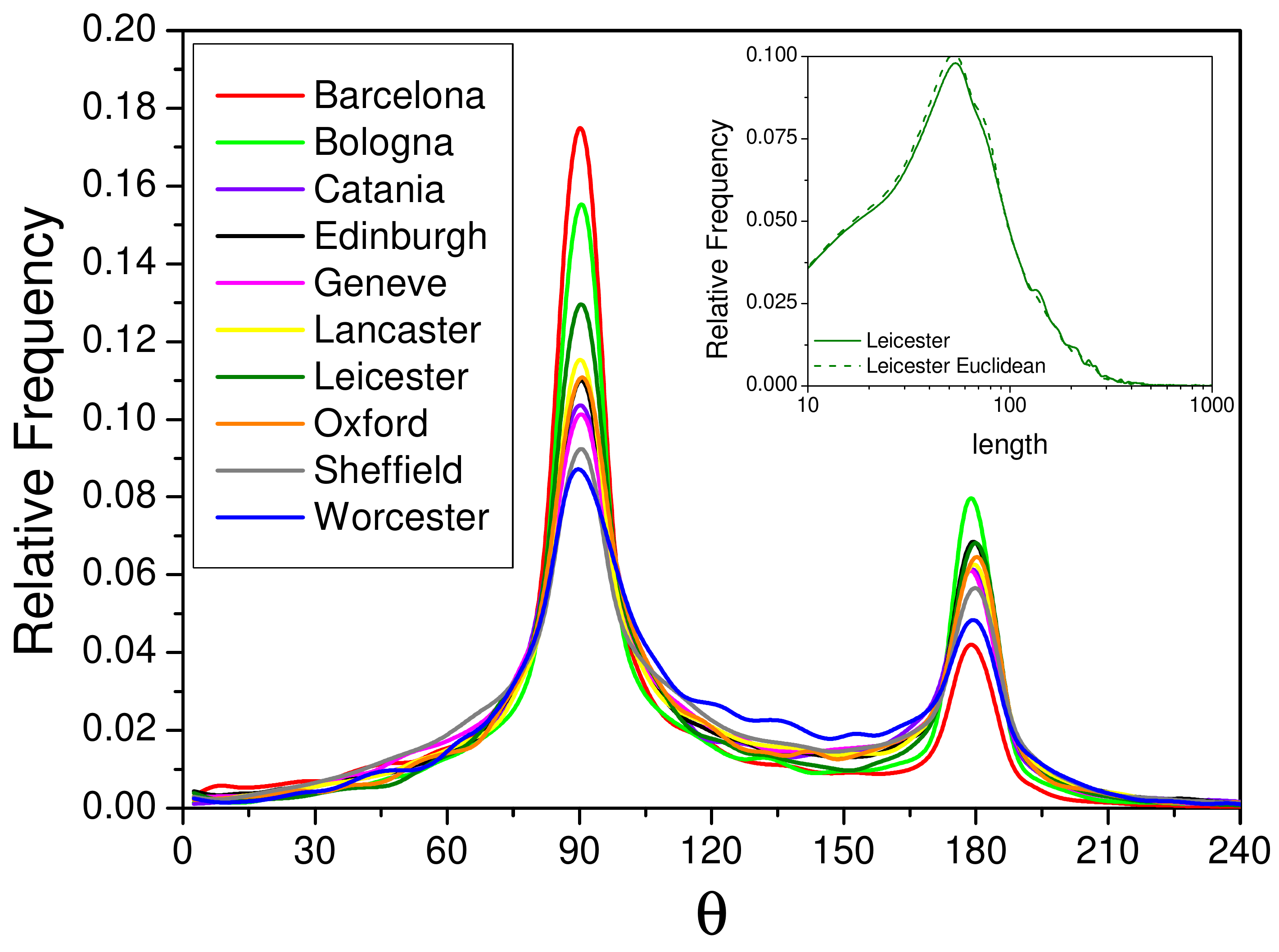}
\caption{Distributions of the angles formed by street intersections of the cities' Euclidean networks. All the cities show the same shape. The inset shows the street length distribution of the original network and the Euclidean one for the city of Leicester.}
\label{f:angles}
\end{figure}

Streets are not always straight lines. In order to study the distribution of angles formed by streets at intersections we must use an equivalent network in which all the streets are represented by straight lines (i.e. substituting the curved streets with straight ones) and where the link weight is equal to the Euclidean distance between its end nodes. We name this network \emph{Euclidean network}. As for all the generalization models, results should be interpreted with caution because of the effect that the approximation may produce on the real structure of the network. Such caution suggests a preliminary test. The inset in Fig.~\ref{f:angles} shows small divergences between the distribution of street length in the original and the Euclidean networks for the city of Leicester. Such simple test confirms that the Euclidean generalization does not lose relevant information, i.e. that streets in cities are not always straight, but predominantly so, confirming the finding provided in \cite{Chan2011}. Therefore we are confident to study the distribution of the angles formed by street intersections in the Euclidean network as a reliable approximation of the original. 
The distributions of the angles formed by streets at intersections are shown in Fig.~\ref{f:angles}. At a glance, we note that all the cities share the same behavior exhibiting a double peak shaped distribution around the characteristic values of 90 and 180 degrees, respectively. This finding confirms the analogous results that have been found in other kinds of spatial planar networks, like those of leaf venations \cite{Couder2002, Bohn2002}, as a result of tensorial stress fields or simple force models.

%

\begin{figure}[ht!]
\centering
\includegraphics [width=7cm] {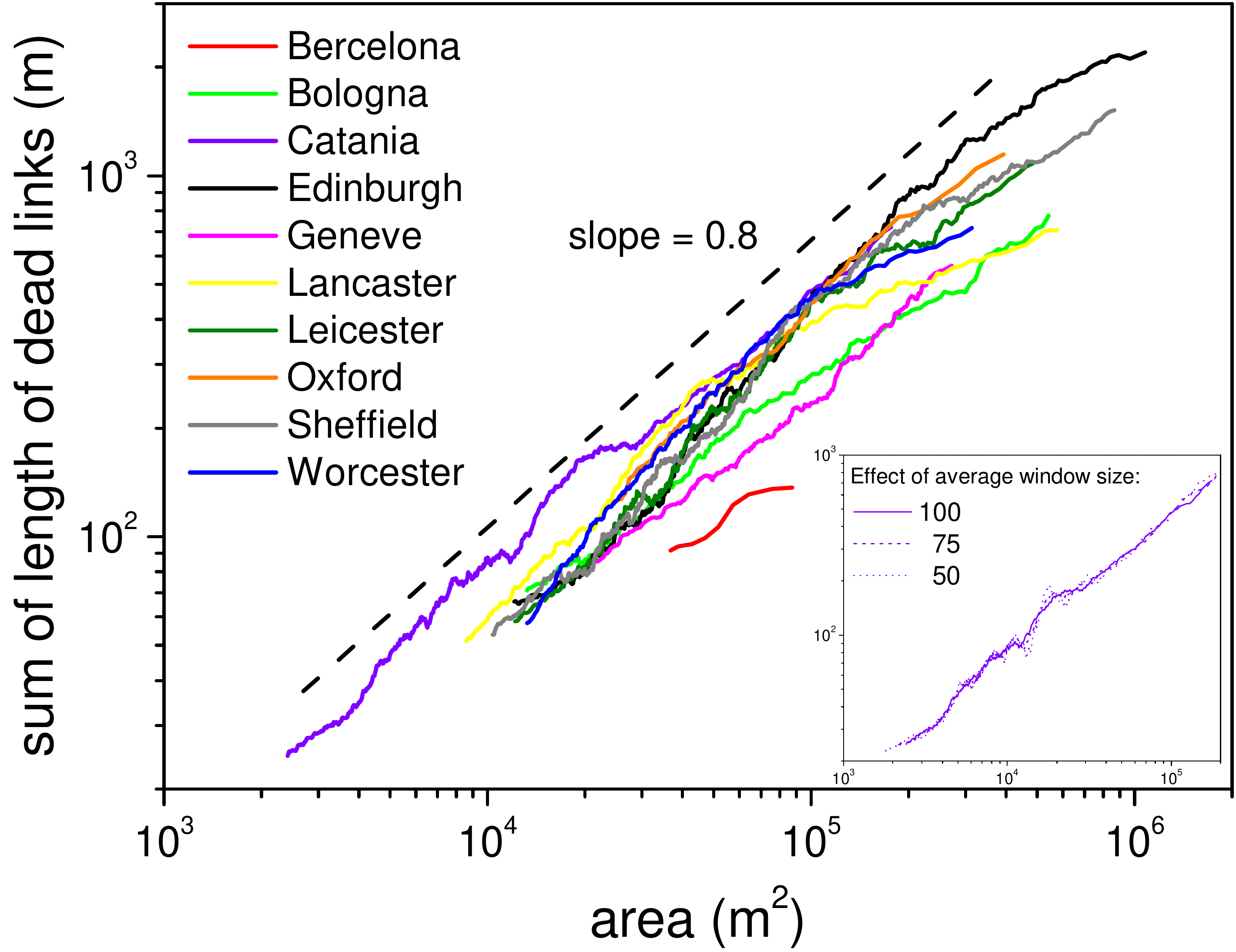}
\caption{Total length of trees like (dead ends) links as a function of the cycle area for the considered cities. The distributions show a power-law behavior with an exponent around 0.8. }
\label{f:deadends}
\end{figure}

The incidence of certain motifs in complex networks is a long standing object of investigation, \cite{Alon2002}. Here we want to focus not on the cycle's shape and quantity, but on the relation between the total length of dead links within a given cycle and its area. Even if we are measuring static systems, i.e. systems that do not change in time, we should remember that a cycle is the result of an evolutionary process that starts with short dead end streets sprouting from the longest edges of the cycles and then extending towards the opposite edge until splitting the original cycle in two smaller sub-cycles. Dead end streets can be interpreted as sprouts of new cycles in parcels still subjected to evolution or as crystallized fractures that do not undergo further development. Their quantity is given by the index $f$ shown in Tab.~\ref{t:ppts} and it can be thought of as an estimator of the abundance of cycles in the intermediate evolutionary stage of their lifetime as suggested in \cite{Barthelemy2008}. Such assumptions are supported by the result shown in Fig.~\ref{f:deadends}, where we report the sum of the length of dead links inside each cycle versus the area of the cycle itself for each city. The distribution shows a clear power-law behavior with a common exponent close to 0.8. It is worth to note that the power-law behavior is not affected by any factor like the fraction of dead ends $f$, or the average degree $\langle k \rangle$. Similar results have appeared in \cite{Lammer2006, Perna2011}. Of course, our findings can be truly confirmed only by investigating the evolution of urban streets in time with the support of empirical data.

%

\section{Centralities and city classification}
\label{s:centralities}

The concept of Centrality has been used for many years in network and social science and, starting from the seminal work by Wasserman \cite{WassermanBook}, there has been a growth of literature both regarding centrality on social networks as well as other kinds of networks.  Depending on the definition, centrality can be understood as meaning proximity between nodes, accessibility from other nodes, or being in a strategic position for connecting couple of nodes. It is clear that from different definitions of centrality, a node actor can be placed at different centrality ranks and that the same node can result with high value for a centrality while yielding weak values for another one. Therefore, for different cities, we can reasonably expect slightly different distributions of centralities. Moreover, we can identify how centralities are mutually dependent and correlated. What makes centralities particularly suited for geographical studies is that they can be visualized and mapped.  We are interested in understanding if and how these ranges of correlations and fluctuation can help to classify cities that share the main network morphology. For example, a grid-like network can be different from a radial one, but only looking at the statistical distribution of centralities might not be enough for a proper classification.  This problem has been already analysed by Crucitti and Porta \cite{Porta2006B} where the statistical distribution of Closeness, Betweenness, Straightness and Information centralities have been analysed for a sample of twenty city parcels of one square mile. Crucitti \etal found significant differences between cities and through a cluster analysis they proposed a classification of different urban patterns. The approach proved to be effective in capturing essential features of urban form as emerging in limited samples selected for their inner morphological consistency, but dealing with entire cities poses the problem of the classification of internally complex objects predominantly composed of different parts each possibly exhibiting different properties. So, the question about the validity of such a procedure on a whole city still need a response. In order to validate if centrality indices can be used in the classification of entire cities, we propose here a clustering method based on the Principal Component Analysis (PCA) made on the distribution of centralities and on their moments. Before discussing this part of the research, we must introduce briefly the adopted centrality indices.

\textbf{Betweenness centrality}, $C^B$, is based on the idea that a 
node is more central when it is traversed by a larger number of the 
shortest paths connecting all couple of nodes in the network. 
More precisely, the betweenness of a node $i$ is defined as in 
\cite{WassermanBook, Freeman1977, Freeman1979}:
\begin{equation}
\label{e:bet}
C^B_i = \dfrac{1}{(N-1)(N-2)} \sum_{\substack{j,k \, \in \, \mathcal{N}\\ i \neq k \, , \, j \neq k}} \dfrac{n_{jk}(i)}{n_{jk}} \,,
\end{equation}
where $n_{jk}$ is the number of shortest paths connecting $j$ and $k$, 
while $n_{jk}(i)$ is the number of  shortest paths connecting $j$ and 
$k$ and passing through $i$.

\textbf{Straightness centrality}, $C^S$, originates from the idea 
that the efficiency in the communication between two nodes $i$ and
$j$ is equal to the inverse of the shortest path length, or geodesic, $d_{ij}$ 
\cite{Latora2001}. In the case of a spatial network embedded 
into a Euclidean space, the straightness centrality of node $i$ is 
defined as:
\begin{equation}
\label{e:str}
C^S_i = \dfrac{1}{(N-1)} \sum_{j \, \in \, \mathcal{N} \, j \neq i} \dfrac{d_{ij}^{\text{Eucl.}}}{d_{ij}} \,,
\end{equation}
where $d^{\text{Eucl}}$ is the Euclidean distance between nodes $i$
and $j$ along a straight line. This measurement captures to which
extent the connecting route between nodes $i$ and $j$, let's say 
between each street junctions, deviates from a virtual straight route. 

The \textbf{Closeness centrality} , $C^C$,of a node $i$ is based 
on the concept of minimum distance, in topological sense, i.e. the 
minimum number of edges traversed to get from $i$ to $j$\\ \cite{Vito2006} and is defined as in \cite{Scott2000, Sabidussi1966}: 
\begin{equation}
\label{e:clo}
C^C_i = \dfrac{1}{L_i} = \dfrac{N-1}{\sum\limits_{j \in G} d_{ij}} \,,
\end{equation}
where $L_i$ is the average distance from $i$ to all the other nodes. 
Closeness centrality is a classical centrality index that has 
been widely used in urban geography and econometrics as well 
as in regional planning, where it gives an idea of the cost that 
spatial distance loads on many different kinds of relationships 
that take place between places, people, activities and markets.

The \textbf{Accessibility} $C^A$ is a measure recently introduced by Traven\c{c}olo and Costa\\ \cite{Travencolo2008}. It has been used for studying the property of very different spatial networks. In the case of urban networks, it has been used to investigate the relationship between subway and road systems \cite{Costa2011}. In addition, the accessibility has been found to be closely related to the borders of networks \cite{Travencolo2009}, in the sense that nodes with low accessibility tend to belong to these borders. The $C^A$ of node $i$ measures the ratio of neighbouring nodes that are effectively reached by an agent randomly navigating the network against the actual number of nodes that belong to the neighborhood. More precisely, $C^A$ takes into account the number of nodes effectively accessed by each node of the network, as well as the probabilities of such accesses. First, we evaluate the transition probability $P_{i,j}(h)$ which describes the probability for an agent leaving from node $i$ to reach node $j$ after $h$ steps along a given type of walk. At each step, the agent located at node $q$, chooses a random neighbor of $q$ and jump to it. These rules define a random walk over the network. When the transition probabilities are very heterogeneous, we have low values of accessibility, meaning that the random walks are biased toward a certain number of nodes, which is, less than the number of nodes which can be reached after $h$ steps.
On the other hand, when the transition probabilities are homogeneous, all nodes which can be reached after $h$ steps are accessed, on average, the same number of times. This case corresponds to the highest values of accessibility. The heterogeneity of the transition probabilities is quantified in terms of the classical concept of entropy, so that the mathematical definition of accessibility, of a node $i$, with respect to the number of steps $h$ is given as

\begin{equation}
C^A_i(h) = \exp\left[ -\sum_{j=1}^N P_{i,j}(h) \log P_{i,j}(h) \right],
\end{equation}
Also, we have considered the transition probability for unitary step $(h=1)$, as:

\begin{equation}
P_{i,j}(1) = P_{i,j} = \frac{w_{i,j}}{\sum_{j=1}^N w_{i,j}},
\end{equation}
where $w_{i,j}$ is the weight of the edge $(i,j)$. In order to take
the geography into account, we considered $w_{i,j}= 1/d_{ij}^{\text{Eucl.}}$.
For disconnected nodes, we assume $d_{ij}^{\text{Eucl.}}=\infty$ such that $w_{i,j}=0$.

In order to investigate if the distribution of centralities can describe the main geographical differences within cities, we use the PCA approach.  This well-established method of multivariate statistics implements a projection of the distribution of objects (in our case, cities) from a higher into a lower dimensional space such that the maximum dispersion of the data is observed at the first new axis (or principal variable), and so forth.  This projection is optimal in the sense of fully decorrelating the original data, therefore removing all correlations between the original measurements describing the objects.  So, since the data dispersion is better described by the first principal axes, the remainder axes can be discarded.  PCA is therefore particularly relevant to the presents studies because: (i) it decorrelates the original measurements; (ii) it provides a projection of the data that maximizes their dispersion (i.e. the differences between the cities); and (iii) it allows the visualization of the distribution of objects (when projected into 2D or 3D spaces).   

The PCA consists in obtaining the covariance matrix of the original data and then extracting its eigenvalues and respective eigenvectors. The eigenvalues can be shown to correspond to the variances along each new axis, and the each respective eigenvector component provides the coefficient of the linear combination of the original measurements used to project the original data into the respective axis. Therefore, the effectiveness of PCA in projecting the data can be inferred by inspecting the eigenvalues in descending order.  For instance, if the two largest eigenvalues account for 75\%  of the overall variance, it can be understood that these two axes are describing the original distribution of points in an effective way, and that the other axes can be overlooked without losing much information.

\begin{figure}[ht!]
\centering
\includegraphics [width=8cm] {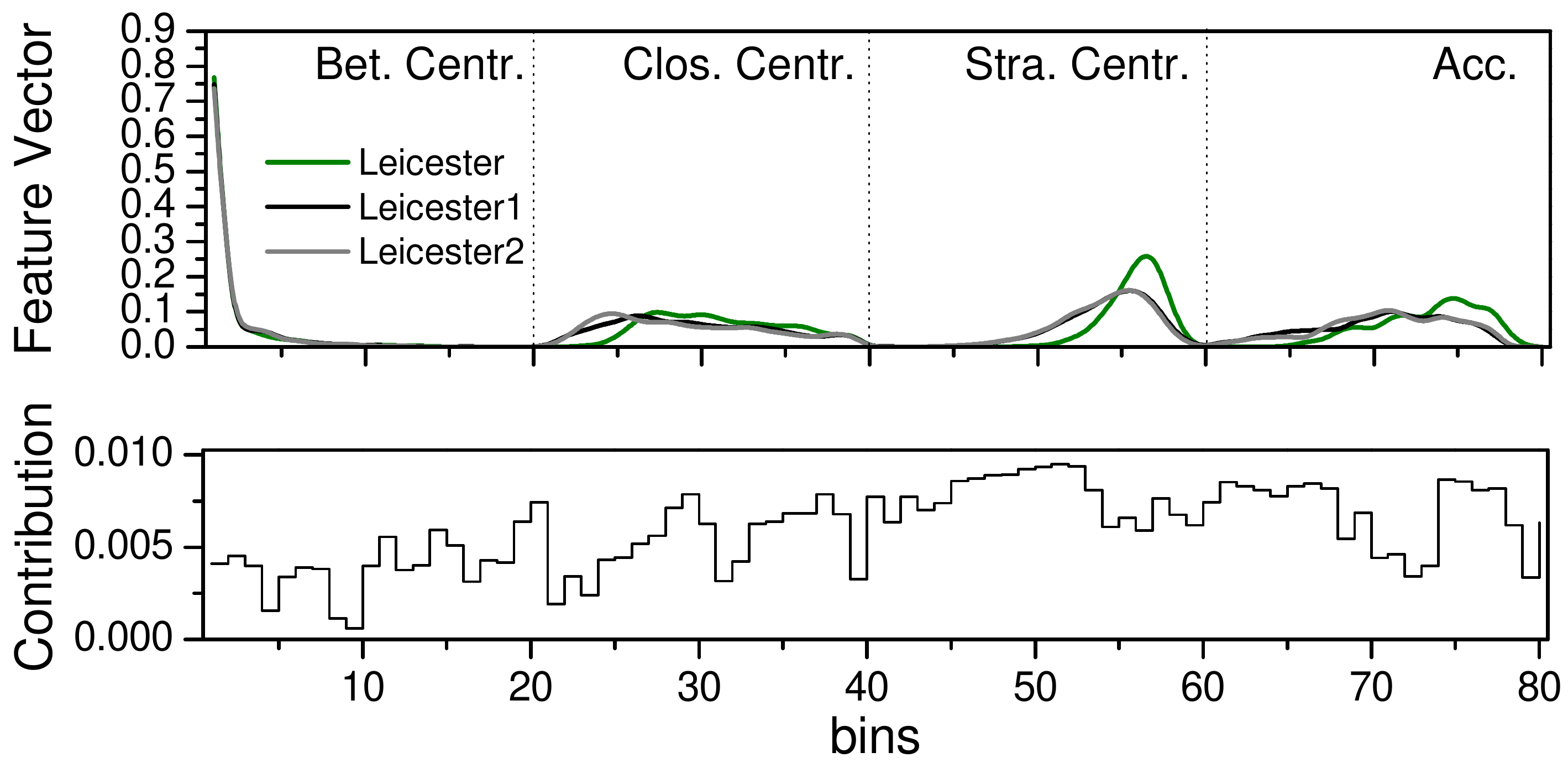}	
\caption{The complete vector centralities and the contribution of each bin to the explanation illustrated in Fig.~\ref{f:pca}}
\label{f:pcacontrib}
\end{figure}

First, we evaluate the histogram of each centrality measurement considering 20 bins. In this way, each dimension corresponds to the relative frequency of centrality values in a small range. The histogram of each one of the four centralities were merger together in order to create an eighty dimension feature vector for each city. For instance, Figure~\ref{f:pcacontrib} (top) shows the feature vector obtained for the city of Leicester.

\begin{figure}[hc!]
\centering
\includegraphics [width=7cm] {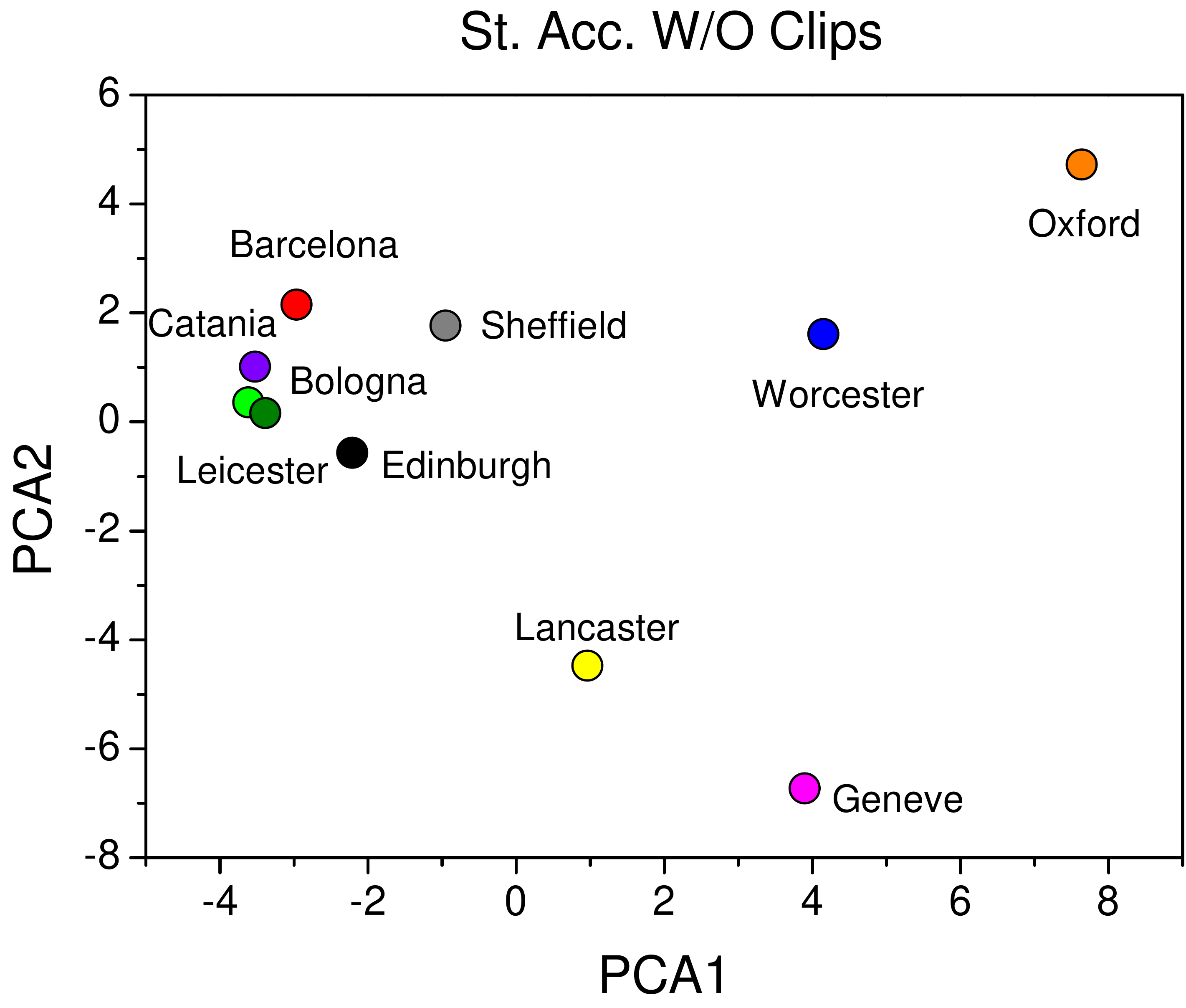}
\caption{PCA using only Straightness and Accessibility}
\label{f:pca}
\end{figure}

\begin{figure}[htp!]
\centering
\includegraphics [width=7cm] {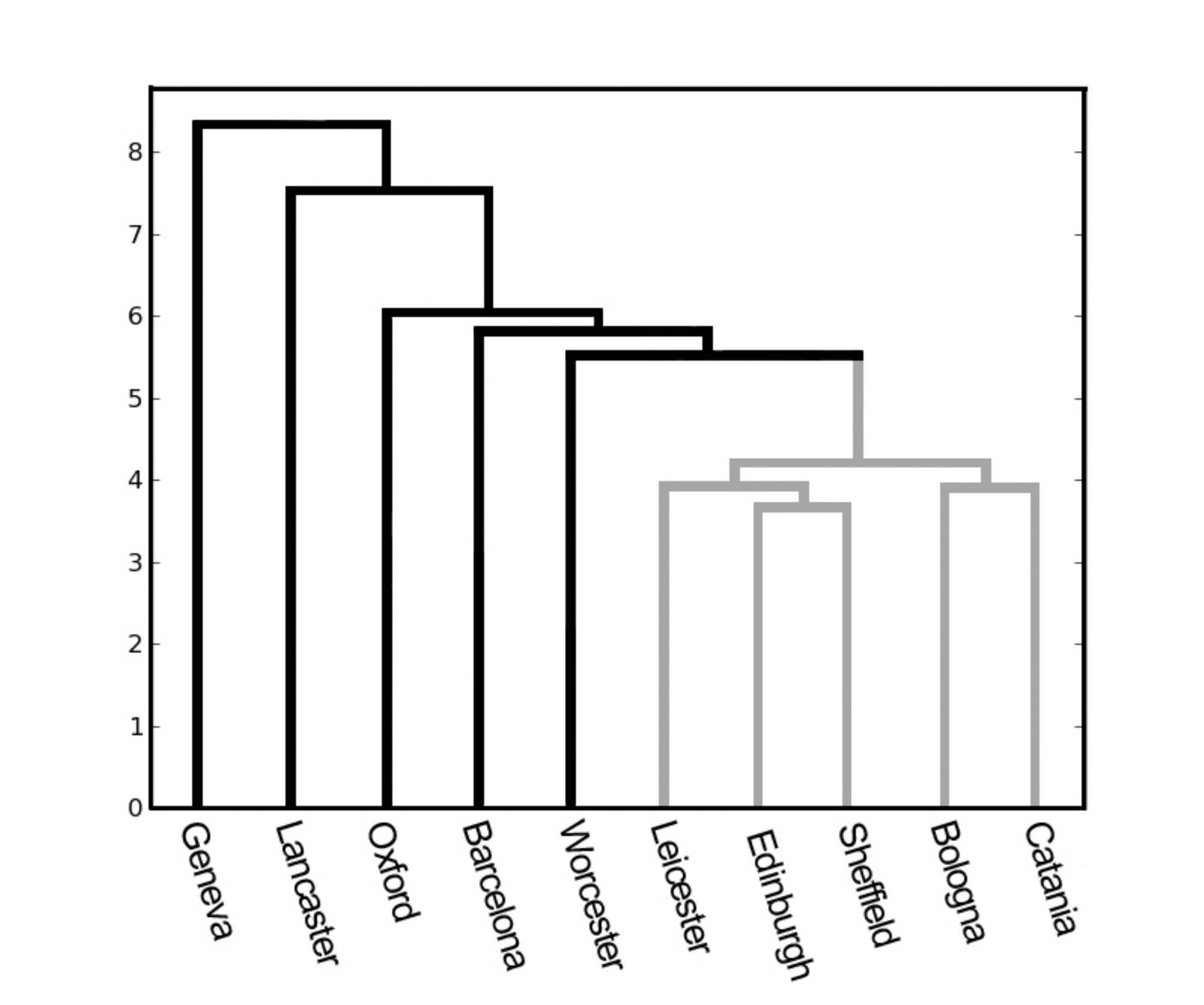}	
\caption{The dendrogram is based on the agglomerative complete 
linkage method on the 12 descriptive components illustrated in 
Fig.8. The dendrogram suggest the existence of two main families 
of cities that are basically with and without big geographical 
constraints. The only exception is given by Barcelona, that even 
without a natural constraint was subjected to a massive and strong
planning operation.}
\label{f:dend}
\end{figure}

In the second approach, the feature vectors were created by considering the 20th first moments of the centralities distributions merged together, also generating an eighty dimension vector. It is important to note that, both first and second approaches provided similar results. Finally, the dimension of the feature vectors were reduced from eighty to two, in which each original dimension has contribution according the values shown in Figure~\ref{f:pcacontrib} (bottom) and Fig.~\ref{f:pcacontrib} (top) shows the explanation of each dimension. By plotting the first two dimensions, it is possible to account for almost  50\% of the dispersion, while a clear differentiation between cities can be seen in Fig.~\ref{f:pca}. Since it is impossible to visualize the high-dimensional data so as to try to recognize the clusters, we used an agglomerative clustering method, the so-called complete linkage method to perform this task. In this method, the Euclidean distance between two clusters is given by the value of the shortest distance between any object belonging to these clusters. The final result is shown in Fig.~\ref{f:dend} as a dendrogram, where the colours correspond to the two clusters identified by using a threshold parameter 0.7 of the maximum distance between two cities.

%

%

This figure clearly shows the emergence of one cluster of cities (grey color) and of a group of cities very different each others but still belonging to a single family (black colour), so we may intepret this agglomeration as the existence of two main different classes or typologies of cities. In this classification a clear group composed by Catania, Bologna, Sheffield, Edinburgh and Leicester is separated from Geneve, Lancaster, Oxford, Worcester and Barcelona. Though the reasons behind this separation may be manifold and it needs further experimental tests, we notice that comparing the plans of those cities and the topography of their terrain it is possible to conclude that physical constraints can be a key factor to classify cities. We have found that cities traversed by rivers or bordered by a lake were separated from those that the growth has never been boundend or divided by physical constraints. The exeption of Barcelona, a city with a major planning events (the Plan Cerd\`a) confirms this hypothesis posing the great urban planning operations in the list of the geographical constraints.

\section{Discussion}
\label{s:discussion}

In this research we analyse the street network of 10 European cities represented in a primal way, where intersections are translated into the nodes of a network and the connecting streets into the links. We study first the geometric properties of the networks and then the way centrality is distributed over them according to four different definition of centrality. We show that selected cities share several universal geometric patterns, such as the average street length (which is limited within a remarkably narrow range of values), the distribution of angles between streets at intersections, and the distribution of the total length of dead end streets as a function of the area of the cycle they belong to. In addition, we confirm that the distribution of street lengths follows universal power law behavior in the long tail of the dataset, i.e. for high values of street length. However, we highlight that the conventional way to represent the distribution of street lengths in search of the power law behavior leaves out of the picture the vast majority of the data, i.e. the number of streets whose length distribution cannot be accurately fitted by a power law function is in the order of the 90\% of the whole dataset. We therefore investigated the actual behaviour of the street length distribution in a semi-log scale finding the emergence of remarkably different behaviours that seemingly reflect the diversity of local history and conditions. Finally, we found that the distribution of the four centrality indices over the street networks allows us to clearly characterize two different clusters of cities that appear to be predominantly informed by major topographic and geographic local features like the presence of rivers or lakes. We also found that planning events of extreme magnitude like the occurrence of the Plan Cerd\`a in Barcelona may account for the uniqueness of this city as measured through the distribution of street centrality.

We show how the analysis of simple geometric properties of the street network as well as a more complex evaluation of the street centrality can highlight differences between cities and how these differences can be used for classifying cities in categories. It is clear that global geometric characteristics of the street network exhibit "universal" behaviors but it is reasonable to argue that, as these behaviors equally emerge in a vast range of transportation networks in nature and technology that like the street networks are mainly planar, the universality of those features may be related to the planarity of the systems rather than a particular "nature" of cities as such. 


Regarding the analysis of the morphology of cities, we believe our study highlights that if the universal rules governing the evolution of planar hierarchical networks are not discriminated, the extreme differences of their inner urban structure may be easily underestimated and, from a qualitative point of view, it may leave local patterns out of the picture. We focused on this discrimination process showing how to operate at global level in order to highlight local patterns where the extreme diversity of our cities emerges and should be accounted for, especially when we move on to the problem of classification, i.e. that of finding a taxonomy for urban types.

\subsection{Acknowledgments}
The authors thanks Shibu Raman, Tim Jones, Mike Jenks and Colin Pooley for providing the data of some of the analyzed cities.
Emanuele Strano is grateful to Marc Barth{\'e}lemy, Salvatore Scellato and Andrea Perna for their suggestions and help and to the Faculty of Engineer at University of Strathclyde for his MRes grant. Luciano is grateful to FAPESP (05/00587-5) and CNPq (301303/06-1 and 573583/2008-0) for financial support. Matheus Viana thanks FAPESP (2010/16310-0) for his PhD grant.
\label{ackn}

\subsection{References}

\newpage


\end{document}